\documentclass[twocolumn,aps,prb,showpacs,preprintnumbers,groupedaddress]{revtex4-1}
\usepackage{amsmath}
\usepackage{amssymb}
\usepackage{dcolumn}
\usepackage{graphicx}
\usepackage{longtable}
\usepackage{bm}
\usepackage{color}
\usepackage{epsfig}

\begin{document}

\title{Half Metal Transition Driven by Doping Effects in Osmium Double Perovskite} 
\author{Madhav P. Ghimire and Xiao Hu}

\affiliation{International Center for Materials Nanoarchitectonics (WPI-MANA), National Institute for Materials Science,Tsukuba 305-0044, Japan}

\date{\today}

\begin{abstract}
    Using the first-principles density functional approach, we investigate Ca$_2$FeOsO$_6$, a material of double perovskite structure synthesized recently. According to the calculations, Ca$_2$FeOsO$_6$ is a ferrimagnetic Mott-insulator influenced by the cooperative effect of spin-orbit coupling (SOC) and Coulomb interactions of Fe-3$d$ and Os-5$d$ electrons, as well as the crystal field. When Fe is replaced with Ni, the system exhibits half metallic (HM) states desirable for spintronic applications.
In [Ca$_2$Fe$_{1-x}$Ni$_x$OsO$_6$]$_2$, HM ferrimagnetism is observed with $\mu_{\rm tot}=2\mu_{\rm B}$ per unit cell for  doping rate $x=0.5$, whereas HM antiferromagnetism (HMAFM) with nearly zero spin magnetization in the unit cell for $x=1$, respectively. It is emphasized that half metallicity is retained even with SOC effect due to the large exchange-splitting between spin-up and spin-down bands close to the Fermi level.
\end{abstract}

\maketitle

\section{Introduction}
Double perovskite oxides have been widely investigated due to their interesting properties useful for spintronic applications \cite{wolf,anderson}. 
With the general formula A$_2$BB$'$O$_6$, A is usually an alkaline earth or rare-earth element, and B and B$'$ are the transition metal elements, both of which are composed of edge-shared octahedra. 
Depending on the choice of B and B$'$ cations, these compounds  show a variety of electrical and magnetic properties, namely metallicity, half metallicity, insulator as well as ferromagnetism, ferrimagnetism, and antiferromagnetism \cite{anderson}. 
The discovery of room-temperature colossal magnetoresistance and half metallicity in Sr$_2$FeTO$_6$ (where T=Mo, Re) \cite {kobayashi,sarma,tokura}, multiferroicity in Bi$_2$NiMnO$_6$ \cite{shimakawa}, magneto-dielectricity in La$_2$NiMnO$_6$ \cite{rogado}, leads to intensive study in double perovskite materials. 
Recent researches have been devoted to understanding the electronic and magnetic properties of hybrid 3$d$-4$d$(5$d$) double perovskites which exhibit high spin polarization \cite{onur,meetei} essential for device applications at room temperature. 

Half metals (HMs) are a class of materials which are metallic in one spin channel, while insulating in the opposite spin channel due to the asymmetric band structure \cite{groot1,xiao1,coey,pickett1,felser1,katsnelson,leuken}. 
HMs  allow spin polarized currents to flow without any external operation, and thus are very useful for spintronics applications. The total spin-moment per unit cell is quantized in units of Bohr magneton ($\mu_{\rm B}$) due to insulating state in one spin-channel. 
HMs have been identified in several groups of materials \cite{pickett2,min1,uehara,mpark,ywang,pardo,xiao2,
akai,xiao3,xiao4,muller,mpg,mpg1,mpg2}.
Most of the experimentally available HMs are either ferromagnets (FM) or ferrimagnets (FiM) which give non-zero integer moments.
It was noted that spin-polarized current can be hampered by stray fields which stabilize magnetic domains \cite{coey,muller}. This drawback can be overcome by HM antiferromagnets (HMAFM), a subclass of HMs characterized further by zero spin magnetization per unit cell \cite{leuken,xiao1}.
A number of materials with DPs structures have been predicted as possible candidates of HMAFMs \cite{min1,pickett2,uehara,mpark,ywang,pardo}.

Because of the recent success in synthesizing  osmium oxides, osmates have been attracting significant interests yielding various unconventional phases.  For instance, unusual superconductivity is observed in A$_2$Os$_2$O$_7$ (A=Cs, Rb and K) \cite{hiroi}. Magnetically driven metal-insulator transition is found in Cd$_2$Os$_2$O$_7$ \cite{jyamaura} and NaOsO$_3$ \cite{calder}, and the ferroelectric-type structural transition has been discovered in metallic LiOsO$_3$ \cite{shi}.  Among the double perovskites,  magnetic insulating states in Sr$_2$MOsO$_6$ (where M=Cu, Ni) \cite{tian}, Mott-insulating ferromagnetic state in Ba$_2$NaOsO$_6$  \cite{jiang}, and half semi-metallic antiferromagnetism in Sr$_2$CrOsO$_6$ \cite{lee} are a few examples that have been reported.
A newly synthesized double perovskite material Ca$_2$FeOsO$_6$   comes into our attention which is reported to be a FiM insulator driven by lattice distortion with high Curie temperature ($T_{\rm c}$)\cite{feng}. The crystal has unique properties suitable for spintronic device applications: Fe and Os atoms carry on opposite spin magnetizations, and octahedra exhibit strong crystal distortion which may induce strong crystal field that helps in splitting the spin-up and spin-down bands. The origin of novel FiM state with $T_{\rm c}$ of 320K opens new possibility of realizing device applications at room temperature. We dope Ni atom having charge state ${+2}$ to replace Fe atom with charge state ${+3}$ to study its influence on the parent material Ca$_2$FeOsO$_6$. 

We have performed first-principles density-functional calculations on Ca$_2$FeOsO$_6$ (CFOO). It is found that CFOO is a FiM Mott insulator with total angular moment $\mu_{\rm tot}=4\mu_{\rm B}$ per unit cell of [Ca$_2$FeOsO$_6$]$_2$.
This material is interesting in the sense that the topmost valence states close to Fermi level ($E_{\rm F}$) are exclusively spin-down bands contributed from $5d$ electrons of Os atoms, and the  element Fe has no influence on the electronic state near $E_{\rm F}$. Therefore, replacing Fe by $3d$ elements with more than five valence electrons one can make a fine control on charge and spin. Specifically, we consider the replacement of Fe atom by the Ni atom  which shows interesting properties desirable in spintronics.

We find that the material [Ca$_2$Fe$_{1-x}$Ni$_x$OsO$_6$]$_2$ is HMFiM with $\mu_{\rm tot}=2\mu_{\rm B}$ at $x=0.5$ and nearly compensated HMAFM with $\mu_{\rm tot}=0.3\mu_{\rm B}$ at $x=1$. 
The interplay among Coulomb repulsion, SOC and the crystal field plays an important role in this material. 

The organization of this paper is as follows: Section~II describes the details on crystal structure and methods. Section~III presents the results on the parent material [Ca$_2$FeOsO$_6$]$_2$, while Sec. IV discusses the results on the possibility of obtaining half metallicity in the doped materials [Ca$_2$Fe$_{1-x}$Ni$_x$OsO$_6$]$_2$. Section~V contains the discussions and the conclusions are drawn in Sec.~VI.
\section{Crystal structures and methods}
 In the double perovskite [Ca$_2$FeOsO$_6$]$_2$ (CFOO), the transition metal Fe and Os occupy B and B$'$ sites in an ordered way. 
The crystal structure of CFOO shown in Fig.~1 falls in the space group \emph{P}2$_1/n$ with monoclinic-distortion. It has structural distortions due to the tilting and rotation of the two corner-sharing FeO$_6$ and OsO$_6$ octahedra.

 \begin{figure}[t]
    \centering
    \psfig{figure=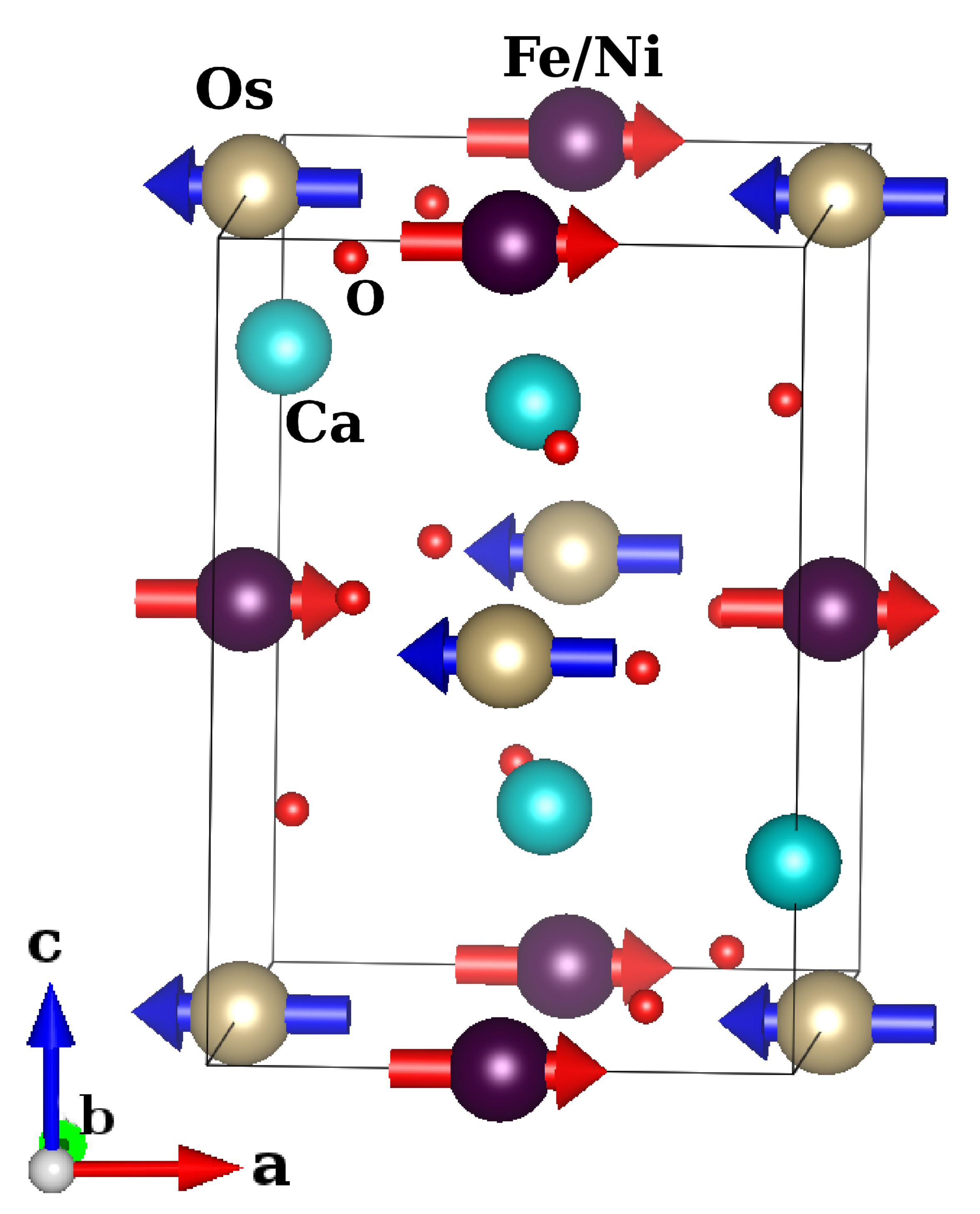,width=2.1in,height=2.4in}
    \caption{Double perovskite structure of [Ca$_2$FeOsO$_6$]$_2$ and [Ca$_2$NiOsO$_6$]$_2$.
    The red (blue) arrows indicate the direction of Fe/Ni(Os) spins along the $a$ direction which is the easy axis.}
    \label{structure}
  \end{figure}

The electronic and magnetic structure calculations were performed within density-functional theory (DFT) by using the full-potential linearized augmented plane wave plus local orbital method implemented in the WIEN2k code \cite{blaha}.
The atomic sphere radii $R_{\rm MT}$ were 2.17, 1.99, 1.99, 2.0 and 1.64 Bohr for Ca, Fe, Ni, Os and O respectively. A set of 1000 $k$-points were used in the full Brillouin zone. The standard generalized-gradient approximation (GGA) exchange-correlation potential within the PBE-scheme \cite{perdew} were used with Coulomb interaction $U$ of 5eV for Fe (Ni) and 1.5eV for Os, respectively \cite{anisimov,note}.
Spin-orbit coupling is considered via a second variational step using the scalar-relativistic eigenfunctions as basis \cite{kunes}.
We checked different magnetic configurations of Fe, Ni and Os sites and found that the FiM structure shown in Fig.~1 is the ground state consistent with the experiments\cite{feng,macquart}. 

\section{{\bf{Parent material 
  [C\lowercase{a}$\mathbf{_{2}}$F\lowercase{e}O\lowercase{s}O$\mathbf{_6}$]$\mathbf{_2}$}}}

   In CFOO, the transition element Fe nominally takes the charge state $+3$ with $3d^5$ configuration. These five valence electrons occupy the $t_{2g}$ and $e_g$ orbits resulting to a high-spin (HS) state of Fe.
On the other hand, Os takes the charge state $+5$ with $5d^3$ configuration, where three valence electrons occupy the $t_{2g}$ orbits giving rise to HS state. 
\begin{figure}[t]
    \centering
    \psfig{figure=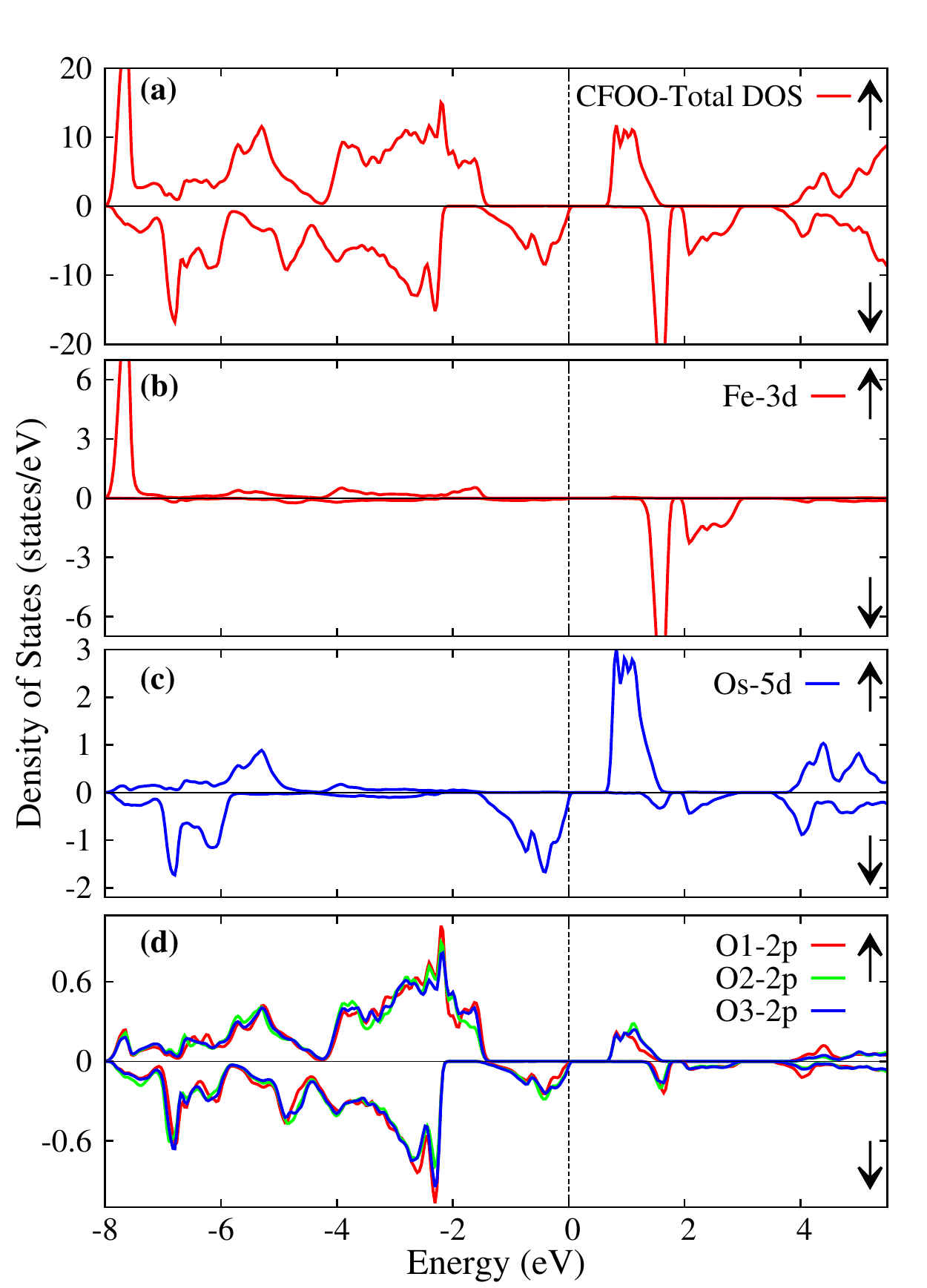,width=2.9in,height=3.7in}
    \caption{Density of states for [Ca$_2$FeOsO$_6$]$_2$ in spin-up ($\uparrow$) and spin-down ($\downarrow$) channels: (a) total, (b) Fe-3$d$ states, (c) Os-5$d$ states and (d) three in-equivalent oxygen-2$p$ states.}
    \label{Fig2}
\end{figure}

To gain insight into the electronic properties of CFOO the spin-resolved total and partial density of states (DOS) in spin-up and spin-down channels are shown in Fig.~2. 
According to first-principles calculations, there is an energy gap of $\sim0.8$eV at $E_{\rm F}$, indicating clearly that CFOO is a Mott insulator. This result is consistent with the recent experimental report of the insulating state in CFOO \cite{feng}.
Fe-$3d$ (i.e., $t_{2g}^3$ and $e_{g}^2$) states in spin-up channel are fully occupied and remain deep in the valence region, whereas the states in spin-down channel lies in the conduction region. 
Unlike Fe, Os-$5d$ (i.e., $t_{2g}^3$) states are located in the conduction region with a broad peak for spin-up channel indicating the empty states, whereas in spin-down channel they occupy the topmost valence bands below $E_{\rm F}$. 
As observed in Fig.~2(d), oxygen bands below $E_{\rm F}$ in both spin-channels hybridize with the Fe-$3d$ and Os-$5d$ states. This is caused due to octahedral distortions, where three sorts of oxygen positions with different Fe(Os)-O bond-lengths appear.

The magnetic property of CFOO is of particular interests. At the ground state obtained from first-principles calculations, Fe couples antiferromagnetically with Os. The calculated total  moment ($\mu_{\rm tot}$) is 4.0$\mu_{\rm B}$ per unit cell (see Table~I).  In an ionic picture, each Fe ion carries moment $+5\mu_{\rm B}$ while Os ion carries $-3\mu_{\rm B}$, giving rise to $\mu_{\rm tot}$ $=2\times(+5\mu_{\rm B})+2\times(-3\mu_{\rm B})=4\mu_{\rm B}$ in [Ca$_2$FeOsO$_6$]$_2$, consistent with the first-principles calculations. 
\begin{figure}[t]
     \centering
    \psfig{figure=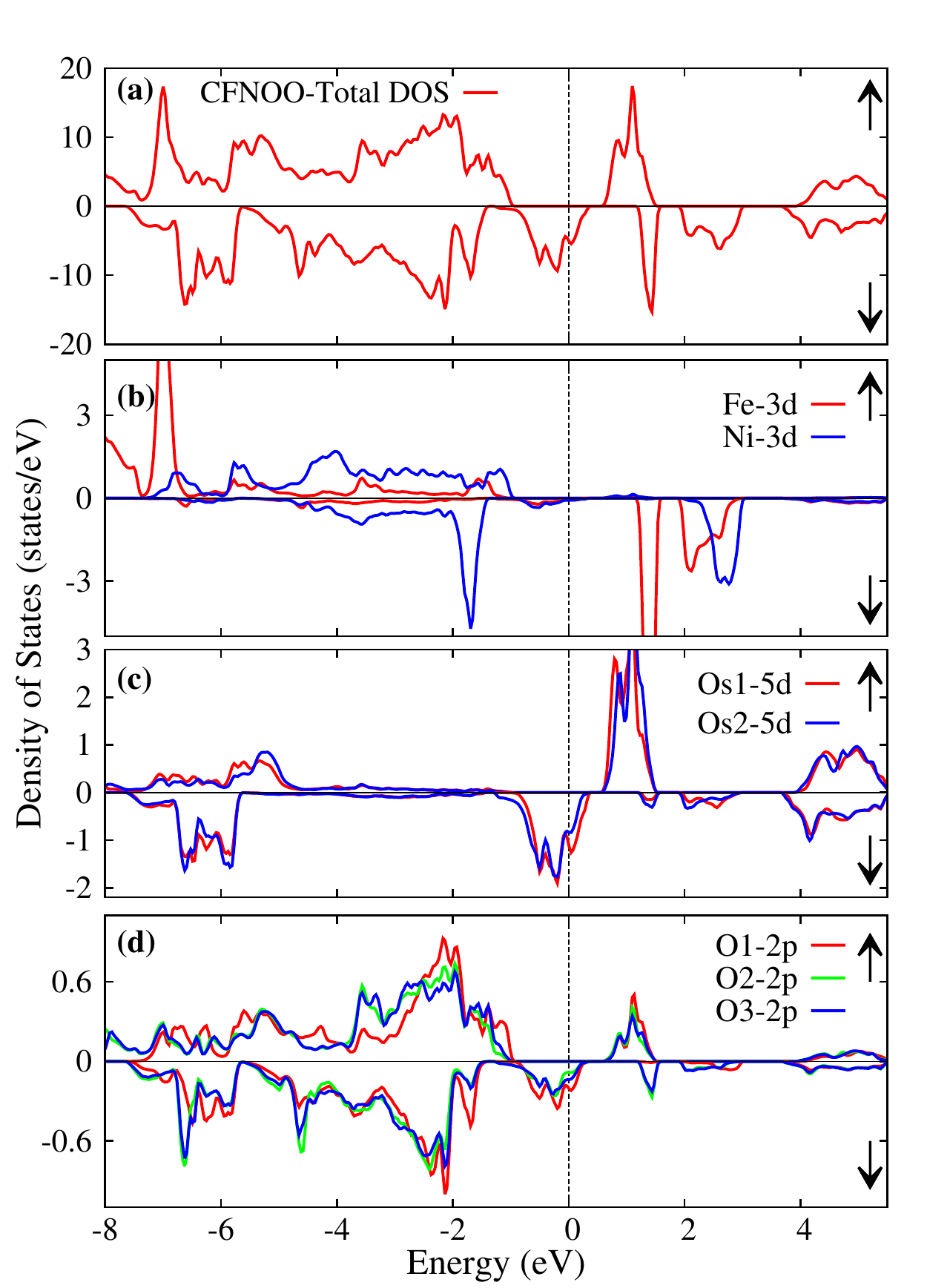,width=2.9in,height=3.7in}
     \caption{Density of states for [Ca$_2$Fe$_{0.5}$Ni$_{0.5}$OsO$_6$]$_2$ in spin-up ($\uparrow$) and spin-down ($\downarrow$) channels: (a) total, (b) Fe-3$d$/Ni-3$d$ states, (c) Os-5$d$ states, and (d) three inequivalent oxygen-2$p$ states}
     \label{Fig3}
\end{figure}

\section{{\bf{Doped materials 
  [C\lowercase{a}$\mathbf{_{2}}$F\lowercase{e}$\mathbf{_{1-x}}$N\lowercase{i}$\mathbf{_x}$O\lowercase{s}O$\mathbf{_6}$]$\mathbf{_2}$}}}
The above properties makes CFOO a promising candidate for exploring possible HM states with fine control on charge and magnetic moments. HM states can be achieved in CFOO by doping $3d$ transition metals having valence electron larger than five. 
To be specific, we consider the B-site modification by replacing one Fe atomd with Ni in [Ca$_2$FeOsO$_6$]$_2$, an element having charge state $+2$ with $3d^8$ configuration. 
Out of eight valence electrons five occupy the $t_{2g}$ and $e_g$ orbits of spin-up channel while the remaining three occupy the $t_{2g}$ orbits in spin-down channel giving rise to a moment of $2\mu_B$ per Ni atom.
Ni doping corresponds to addition of three extra electrons to the system. 
Since the charge states of Fe($+3$) and Ni($+2$) differ from each other, the Os atom interacting with Fe has a charge state $+5$ and $5d^3$ configuration while the other one interacting with Ni has $+6$ charge state and $5d^2$ configuration, to maintain charge neutral in the system.
Thus, Ni doping will influence the system by (i) shifting  the Os bands towards the conduction region, and (ii) compensating magnetic moment by $2\mu_B$ per each Fe replacement.
In this way, one can modify the material to design the desired HMFiM and HMAFM.
\begin{figure}[t]
    \centering
     \psfig{figure=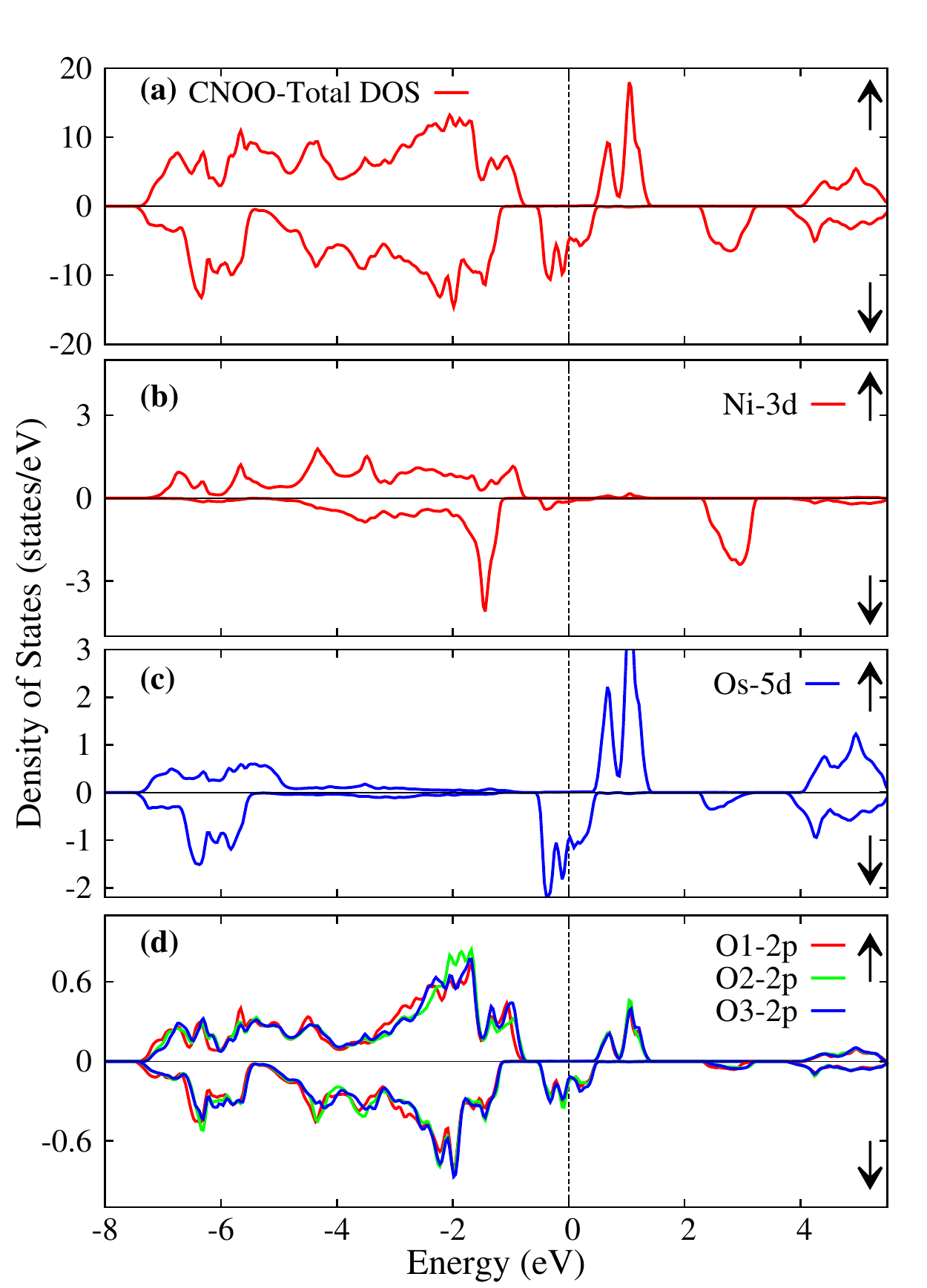,width=2.9in,height=3.7in}
    \caption{Density of states for [Ca$_2$NiOsO$_6$]$_2$ in spin-up ($\uparrow$) and spin-down ($\downarrow$) channels: (a) total, (b) Ni-3$d$ states, (c) Os-5$d$ states and (d) three in-equivalent oxygen-2$p$ states.}
    \label{Fig4}
\end{figure}

\begin{figure*}[bth]
    \centering
     \psfig{figure=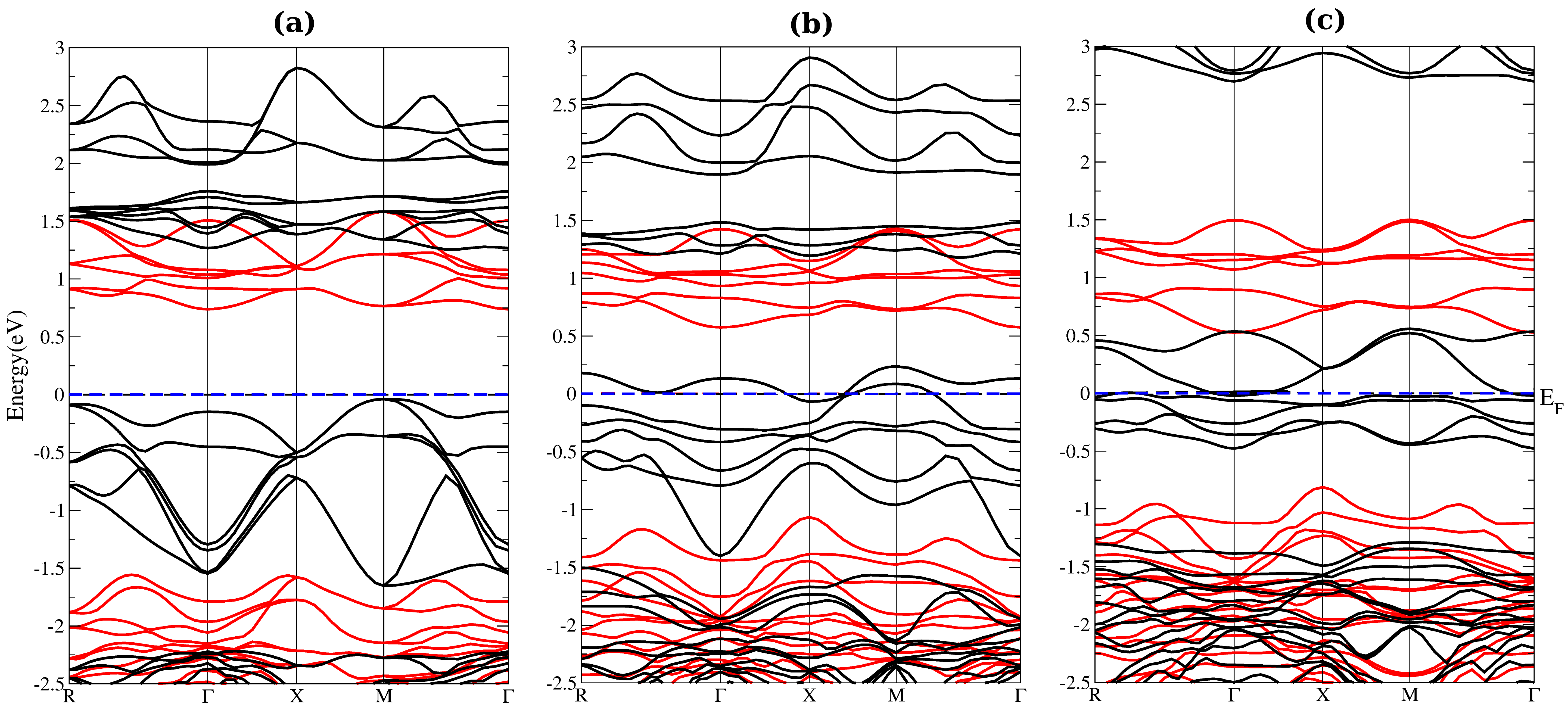,width=7.in,height=3.in}
    \caption{Band structures for doping rate (a) $x=0$, (b) $x=0.5$ and (c) $x=1$ in [Ca$_2$Fe$_{1-x}$Ni$_x$OsO$_6$]$_2$ for spin-up (red) and spin-down (black) channels.}
    \label{Fig5}
\end{figure*}

We perform first-principles calculations to check the above idea for replacement rate $x=0.5$ in [Ca$_2$Fe$_{1-x}$Ni$_x$OsO$_6$]$_2$. 
As shown in Fig.~3, Fe-$3d$ states in spin-up channel are fully occupied lying deep in the valence region, whereas for spin-down channel they remain in the conduction region. The Ni-$3d$ bands are also occupied in spin-up channel while the spin-down bands are partially occupied with a localized peak at -1.8eV below $E_{\rm F}$. The remaining bands appear in the conduction region. This happens because five out of eight $d$ electrons from Ni occupy the spin-up channel and the three remaining electrons go to spin-down channel to fill the $t_{2g}$ orbits. 
Due to repulsive interaction from spin-down Ni-$3d$ electrons, Os-$t_{2g}$ electrons which were originally lying at the topmost valence region in the parent material (see Fig.~2) shift towards the conduction region crossing $E_{\rm F}$ (see Fig.~5(b)). As the results, the Os-$t_{2g}$ states crossing $E_{\rm F}$ form a continuous band and give rise to metallic state for spin-down channel. 
Hence, with spin-up channel insulating and spin-down channel metallic, the material is a half metal.

Let us now focus to the most interesting case of replacement rate $x=1$, where both Fe atoms are replaced by Ni atoms within unit-cell. From the first principles calculations the FiM ground state is $\sim135$meV less than the first excited AFM state. Due to the presence of Ni atom alone on the B-site, Os atom fully attains the charge state $+6$ with $5d^2$ configuration unlike in the parent material [Ca$_2$FeOsO$_6$]$_2$.
For Ni of $d^8$, three $t_{2g}$ and two $e_g$ orbits in spin-up channel and three $t_{2g}$ orbits in spin-down channel are occupied. For Os of $d^2$, only two $t_{2g}$ orbits in spin-down channel are occupied. The system is then expected to favor half metallicity with full compensation of total moment. 
As revealed by the DOS shown in Fig.~4, Ni-$3d$ states are fully occupied in spin-up channel by five electrons and the remaining electrons go to occupy the $t_{2g}$ orbits in spin-down channel. This is evident with the occupation of Ni-3$d$ states in the valence region with a localized peak at -1.6eV below $E_{\rm F}$. The unoccupied $e_g$ band appears far in the conduction region. 
The Os-5$d$ states are almost empty in spin-up channel but has partial occupation with a band crossing $E_{\rm F}$ in spin-down channel. 
This is caused mainly by the spin-down electrons of Ni which push the Os-$t_{2g}$ states up towards the conduction region. Thus, two of the $t_{2g}$ states of Os atom remain in the valence region while the unoccupied state moves to the conduction region crossing $E_{\rm F}$ (see also Fig.~5(c)). 
Due to crystal distortion, gap could not open-up among the $t_{2g}$ states of Os atom which results in metallic state in spin-down channel. 
Oxygen 2$p$ states are found to hybridize strongly with Os-5$d$ states near $E_{\rm F}$ in both spin channels.
Thus, with spin-up channel insulating and spin-down channel metallic, the system turns to a HM as clearly seen in Fig.~4.

As summarized in Table~I, two replaced Fe atoms take away $\mu\simeq 2\mu_{\rm B}$, and the charge transfer in two Os atoms associated on doping reduces $\mu\simeq 2\mu_{\rm B}$ further,  resulting in compensation of total moment to zero.
The results obtained by first-principles calculations are consistent with the ionic picture.
These features can also be seen from the spin-density isosurface plot in Fig.~5(b). 
 With the zero total moment and HM property, the material [Ca$_2$NiOsO$_6$]$_2$
should be a HMAFM. 
However, SOC induces an orbital moment of $\sim0.17\mu_{\rm B}$ across Os resulting in the total moment of 0.3$\mu_{\rm B}$ per unit cell. Hence the material may be called a nearly compensated half metal.

 \begin{table}[t]
  \centering
  \caption{Moments per atom of Fe[Ni] and Os, one set of three in-equivalent oxygen atoms and unit cell ($\mu_{\rm tot}$)
   for replacement rate $x$ in [Ca$_2$Fe$_{1-x}$Ni$_x$OsO$_6$]$_2$ from first-principles calculations. The unit of moments is the Bohr magneton $\mu_{\rm B}$. The contributions from individual atoms are within muffin-tins while the total angular moment includes those from interstitial regime.}
  \begin{ruledtabular}
  \begin{tabular}{r c c c c }
{ $x$} &    Fe [Ni]    &  Os    &  O &  {$\mu_{\rm tot}$} \\\cline{1-5}
     0  & 4.13   & -1.6  & -0.11   & 4.0 \\
   0.5  & 4.13 [1.68]   & -1.36   & -0.15   & 2.0 \\
   1  & [1.68]   & -1.09   & -0.14   & 0.3 \\
  \end{tabular}
  \end{ruledtabular}
  \label{tab:1}
 \end{table}

 \begin{figure}[t]
     \centering
     \psfig{figure=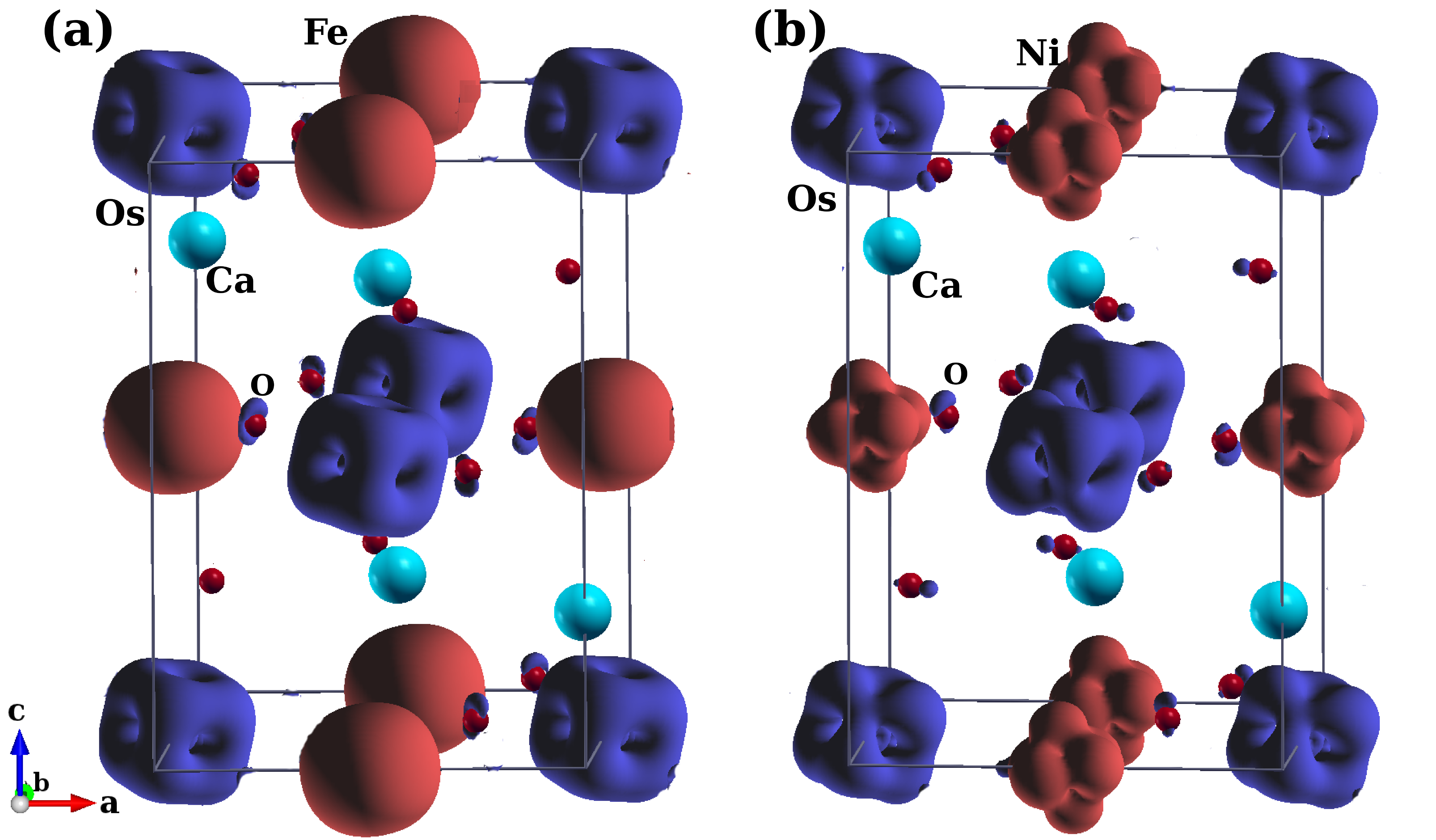,width=3.2in,height=2.in}    
     \caption{Isosurface of spin magnetization density at $\pm$0.21 $e/$\AA$^3$ with red (blue) for spin up (down): (a) parent material [Ca$_2$FeOsO$_6$]$_2$, (b) material with full Fe replacement [Ca$_2$NiOsO$_6$]$_2$.}
     \label{Fig6}
 \end{figure}

\section{Discussions}
     In order to give a clear picture on how doping changes the electronic structure for $x=0$, 0.5 and 1 in [Ca$_2$Fe$_{1-x}$Ni$_x$OsO$_6$]$_2$, band structure plots are shown in Fig.~5. It should be noted that six bands lying in the topmost valence region below $E_{\rm F}$ in the parent material (i.e. $x=0$) are the $t_{2g}$ bands contributed by two Os atoms in spin-down channel (see Fig.~5(a)). It is interesting to note that half replacement of Fe by Ni corresponds to shifting of one of the $t_{2g}$ band towards the conduction region. As clearly observed for $x=0.5$, five out of six of the $t_{2g}$ bands are occupied while the remaining one band moves to the conduction region. As similar case is observed for $x=1$ where four $t_{2g}$ bands remain below $E_{\rm F}$ while the two bands go to conduction region. 
    
    In the present materials, SOC is crucially important due to the presence of heavy elements such as Os. The orbital moments obtained from first-principles calculations for Os ($0.1\sim0.2\mu_{\rm B}$) are in accordance with the Hund's third rule \cite{kittel}. SOC reduces the band gap by $\sim0.1$eV through the broadening of Os-$5d$ states in the parent material. 
    SOC induces the total moment of $0.3\mu_{\rm B}$ per unit cell for [Ca$_2$NiOsO$_6$]$_2$ which is in reasonable agreement with the experiment \cite{macquart}. 
  
Charge-transfer effect \cite{zaanen} is prominent between Os and oxygen via $t_{2g}$ and $p$ states in the parent material. 
Therefore, oxygens get spin-polarized in parallel with the Os ions, consistent with the isosurface plot shown in Fig.~6(a).
The isosurface of Fe-$3d$ states are spherical due to occupation of all the $d$ orbitals in spin-up channel. Os on the other hand shows the $t_{2g}$ characters.
The full replacement of Fe with Ni shows an active $e_g$ like orbitals (see Fig.~6(b)). The size of the Os-$t_{2g}$ isosurface reduces in the doped material due to one electron less than in the parent material.

First-principles calculations on magnetic anisotropy energy indicates $a$ axis of the crystal as the easy axis (see Fig.~1) with anisotropy energy of $\sim 2$meV  and $\sim 4$meV per unit cell for Ca$_2$FeOsO$_6$ and Ca$_2$NiOsO$_6$ respectively. 

Robustness of half metallicity is checked for doping rate $x=1$ by  considering the (i) antisite disorder and (ii) surface effects with a vacuum of 20$\AA$ along 001. We have confirmed that the HM state remains stable in both configurations.

In the present work, HMAFM and HMFiM have been derived from the same parent material. Thus, using them in an integrated system, one can construct a useful device for spintronics applications without suffering from the problem of lattice mismatching.

\section{Conclusions}
Based on the first-principles density functional approach, we propose material tailoring on a Mott insulator [Ca$_2$FeOsO$_6$]$_2$ with double perovskite structure exploiting the cooperative effect from Coulomb interaction, spin-orbit coupling and the crystal field. It is demonstrated that replacing Fe by Ni, one can achieve several half metals. Especially, [Ca$_2$NiOsO$_6$]$_2$ is found to be a nearly compensated half metals, which is ideal for spintronic applications. It is emphasized that the large exchange splitting between spin-up and spin-down bands at the Fermi level retains the half metallicity even in presence of strong spin-orbit coupling.

\acknowledgments
    The authors thank K. Yamaura for providing the crystal information of the parent material. This work was supported by WPI Initiative on Materials Nanoarchitectonics, MEXT, Japan.


\begin{thebibliography}{99}
\bibitem{note}
	 The obtained results for the present materials are robust with $U=2-8$eV for Fe(Ni) and $0.5-2.75$eV  for Os, respectively \cite{anisimov}.
\end{thebibliography}
\end{document}